\documentclass{aa}
\usepackage{graphicx}
\usepackage{txfonts}
\usepackage{natbib}
\bibpunct{(}{)}{;}{a}{}{,}
\usepackage{url}

\newcommand{\nirvana}{{\sc NIRVANA}}
\newcommand{\flash}{{\sc FLASH}}

\newcommand{\mvect}[1]{\mathbf{#1}}
\newcommand{\ud}{\mathrm{d}}
\newcommand{\grad}[1]{\nabla #1}

\usepackage[usenames]{color}

\begin{document}
   \title{Vortex generation in protoplanetary disks with an embedded giant planet}

   \author{M. de Val-Borro
          \inst{1,2}
          \and
          P. Artymowicz
          \inst{3,2}
          \and
          G. D'Angelo
          \inst{4}
          \and
          A. Peplinski
          \inst{2}
          }

   \offprints{M. de Val-Borro\\
        \email{mdeval@cfa.harvard.edu}}

   \institute{
        Harvard-Smithsonian Center for Astrophysics, 60 Garden St., Cambridge, MA 02138, USA
         \and
        Stockholm University, AlbaNova University Center, SE-106 91, Stockholm, Sweden
         \and
         University of Toronto at Scarborough, 1265 Military Trail, Toronto, Ontario M1C 1A4, Canada
         \and
NASA-ARC, Space Science and Astrobiology Division, MS 245-3, Moffett Field, CA 94035, USA}

\titlerunning{Vortex formation in protoplanetary disks}
\authorrunning{M. de Val-Borro et al.}
 
  \date{Received 10 January 2007 / Accepted March 2007}

  \abstract
{Vortices in protoplanetary disks can capture solid
particles and form planetary cores within shorter
timescales than those involved in the standard core-accretion model.} 
{We investigate vortex generation in thin unmagnetized
protoplanetary disks with an embedded giant planet with
planet to star mass ratio
$10^{-4}$ and $10^{-3}$.}
{Two-dimensional hydrodynamical simulations of a
protoplanetary disk with a planet are performed using two
different numerical methods.  The results of the non-linear
simulations are compared with a time-resolved modal analysis
of the azimuthally averaged surface density profiles using
linear perturbation theory.}
{Finite-difference methods implemented in polar coordinates
generate vortices moving along the gap created by Neptune-mass to
Jupiter-mass planets. The modal analysis shows that unstable
modes are generated with growth rate of order $0.3
\Omega_{\mathrm{K}}$ for azimuthal numbers $m=4,5,6$,
where $\Omega_{\mathrm{K}}$ is the local Keplerian
frequency.  Shock-capturing Cartesian-grid codes do not
generate very much vorticity around a giant planet in a
standard protoplanetary disk.  Modal calculations confirm
that the obtained radial profiles of density are less
susceptible to the growth of linear modes on timescales of
several hundreds of orbital periods.
Navier-Stokes viscosity of the order $\nu=10^{-5}$
(in units of $a^2 \Omega_{\mathrm{p}}$) is found
to have a stabilizing effect and prevents the formation
of vortices.
This result holds at
high resolution runs and using different types of boundary
conditions.}
{Giant protoplanets of Neptune-mass to Jupiter-mass can excite
the Rossby wave instability and generate vortices in thin
disks.  The presence of vortices in protoplanetary disks
has implications for planet formation, orbital migration,
and angular momentum transport in disks.}


   \maketitle

\section{Introduction}

Stability of rotationally supported gas disks is an area of active research,
motivated among other reasons, by a need to understand the origin and stability
of hydrodynamics turbulence underlying the so-called anomalous viscosity in
accretion disks.
The concept of $\alpha$-turbulence in accretion disks was introduced
more than three decades ago by \citet{1973A&A....24..337S} to
account for the angular momentum transfer and explain accretion
onto the central object.
The magnetorotational instability (MRI)
has been proposed to explain
the enhanced viscosity
in hot and sufficiently ionized accretion disks
with a Keplerian angular velocity profile
threaded by a weak magnetic field
\citep{1991ApJ...376..214B,1996ApJ...467...76B,1998RvMP...70....1B}.
However, in the context of cold protoplanetary disks, the ionization
by cosmic rays and stellar radiation is limited to the surface layers
of the disk while the so called ``dead zone'' in the vicinity of the
central plane is expected to have low ionization
\citep{1996ApJ...457..355G}.
In some astrophysical systems such as
cataclysmic variables and outer regions of active
galactic nuclei the coupling between the magnetic field and
the gas is also weak and
MHD effects may be negligible.

The stability
of differentially rotating disks
has been considered analytically
and numerically
in the purely hydrodynamical case
\citep{1984MNRAS.208..721P,
1985MNRAS.213..799P,
1986MNRAS.221..339G,
1987MNRAS.225..267P}
with applications to circumstellar disks and
galactic disks.
A rotating isentropic torus with a gradient of
specific angular momentum is found
to be unstable to low-order non-axisymmetric perturbations
due to the Papaloizou-Pringle instability.
Several mechanisms have been proposed that are able to sustain
purely hydrodynamical
turbulence and generate an anomalous $\alpha$-viscosity
in accretion disks
\citep{2000ApJ...533.1023L,
2003ApJ...582..869K,
2005ApJ...629..383M}.
\citet{2005A&A...429....1D} studied non-axisymmetric
instabilities in stratified Keplerian disks using numerical
and analytical methods. A linear instability
appears for Reynolds numbers of order $10^3$
and perturbations with characteristic
scales smaller than the vertical scale of the disk,
assuming the angular velocity
decreases with radius.
These results suggest that despite the stabilizing
effect of the Coriolis force, a Keplerian flow
may undergo a transition to turbulence.
Nevertheless, some of those mechanisms may depend on boundary
or edge effects.

Rossby waves in thin Keplerian disks
have been studied
in the linear approximation
\citep{1999ApJ...513..805L,2000ApJ...533.1023L}
and with fully non-linear numerical simulations
\citep{2001A&A...380..750T}.
The existence of unstable modes has been found to be associated
with radial gradients of an
entropy-modified version of vortensity.
Rossby waves in disks
break up forming vortices in the nonlinear
limit \citep{2001ApJ...551..874L} in agreement
with the predictions from linear theory.
The dispersion relation of this
Rossby Wave Instability (RWI)
is analogous
to the one for Rossby waves in planetary atmospheres.
Reynolds stresses produced by the RWI
can yield outward transport of angular momentum
and accretion onto the central star.
\citet{2006A&A...446L..13V} have discussed the generation of Rossby waves
in the ``dead zones'' of protoplanetary disks where they may
enhance the accretion rate of solids and favor planet formation.
The angular momentum transport in disks around supermassive
black holes at the center of galaxies can also be explained
by the formation of Rossby vortices when there is
a steep enough density gradient
\citep{2003ApJ...598L...7C}. The angular momentum transfer
due to vortices in galactic disks is found to be greater
than in an $\alpha$-viscosity disk.
Recently, \cite{2006ApJ...636L..33T} have described a
magnetohydrodynamics version of the RWI 
and applied it to the study of
the quasiperiodic oscillations in Sgr A*.
Rossby waves can also appear in
thin planetary atmospheres with solid rotation
leading to the formation of vortices like
Jupiter's ``Great red spot'' \citep{1988Natur.331..693M}.

Long-lived vortices are able to capture solid materials
to form massive bodies
and speed up the formation of planetary cores
\citep{1995A&A...295L...1B,1999PhFl...11.2280B,2006ApJ...639..432K}.
The stability of three-dimensional vortices in a
three dimensional stratified disk has been studied by
\citet{2005ApJ...623.1157B,2006JCoPh.219...21B}
using spectral anelastic hydrodynamics
simulations.
They find that vortices are hydrodynamically stable
for several orbits
away from the mid-plane of the disk.
The formation of vortices
in the corotation region
excited by a protoplanet
has been studied
numerically 
by \citet{2001MNRAS.326..833B}
including the
saturation of the corotation torque
and the effects of dissipation
in the non-linear dynamics of the flow.

Several numerical schemes studied by \citet{2006MNRAS.370..529D}
show vortex formation
but the aim of the study did not include
a detailed investigation of vortex generation.
In this paper, we intend to look at this formation process in more
detail. We also wish to check that codes that do not predict vortex
generation do not artificially damp unstable
modes due to the numerical viscosity.
Some simulations produce waves
and vortices at the edge of the gap, which could in principle
interact with the wake to cause semi-periodic disturbances
propagating away along the shock.
In simulations using the Piecewise Parabolic Method (hereafter PPM),
low-$m$ perturbations are observed at the edges of the gap
\citep{2000AN....321..171C}.
Wave-like disturbances with mode number $\sim 5$
are observed at the edge of the gap created
by a Jupiter-size planet
in the numerical results presented by
\citet{2003ApJ...589..556N},
Instabilities close to the planet or along
the edges of the gap created
by a giant planet, as well as the
time variability of the flow near the Roche lobe
may affect the speed and direction of
planetary migration.

In this paper, we study the effect of an annular gap cleared by a
planet on the stability of a protoplanetary disk.
We consider non-axisymmetric linear perturbations to the
inviscid and compressible Euler equations.
In Section~2, we present the semi analytical methods used to
study the stability of disks.
We describe the numerical codes in Section~3.
In Section~4 we present the results of the
numerical simulations and the
perturbative linear analysis.
We discuss the results in the context
of protoplanetary disks in Section~5.
Finally, the numerical diffusivity in our
numerical codes is calibrated
in Appendix~\ref{ap:viscosity}.

\section{Modal analysis}

We perform a modal analysis of analytical
and numerically obtained density profiles,
in order to see if there is agreement
between the vortex generation in the simulations
and growing unstable modes in the linear
stability analysis.
Linear perturbative analysis provides a valuable
tool to study the stability of disks.
The solution of the linearized Euler equations is
treated along the lines of the work of
\citet{1999ApJ...513..805L} and \citet{2000ApJ...533.1023L},
where the stability can be evaluated solving a numerical
eigenvalue problem for a given profile.
Alternatively, the growth of the initial perturbations
can be determined by solving the equations
as an initial value problem.

We consider non-axisymmetric
small perturbations
sinusoidally varying in azimuth to the inviscid Euler equations
\begin{eqnarray}
\frac{\partial\tilde{\Sigma}}{\partial t} + \nabla \cdot (\tilde{\Sigma} \tilde{\mvect{v}}) &=& 0 \\
\frac{\partial\tilde{\mvect{v}}}{\partial t} + (\tilde{\mvect{v}} \cdot \nabla ) \tilde{\mvect{v}} & = & - \frac{1}{\tilde{\Sigma}}\grad{\tilde{P}} - \grad{\Phi}
\label{eq:Euler}
\end{eqnarray}
where $\tilde{\Sigma}$ is the surface mass density,
$\tilde{P}$ the vertically integrated pressure,
$\tilde{\mvect{v}}$ the velocity of the flow and
$\Phi$ is the gravitational potential.
The perturbed quantities 
have the form
$\tilde{\Sigma} = \Sigma + \delta \Sigma(r,\phi,t)$,
$\tilde{P} = P + \delta P(r,\phi,t)$ and
$\tilde{\mvect{v}} = {\mvect{v}} + \delta {\mvect{v}}(r,\phi,t)$ 
with perturbations in the disk plane depending on the azimuthal angle as
$\propto \exp \left[i(m\phi - \omega t)\right]$,
where $m$ is the azimuthal mode number 
and $\omega=\omega_r + i \gamma$ is the complex mode frequency.
The initial velocity component in the radial direction
is neglected and the angular velocity $\Omega$ is
obtained from the force balance between gravity,
pressure gradients and centrifugal force
\begin{equation}
\label{eq:omega}
\Omega^2 = \frac{1}{r} \left[ 
\frac{1}{\Sigma} \frac{\ud P}{\ud r}  + 
\frac{\ud \Phi}{\ud r} \right]
\end{equation}
We use a barotropic equation of state, $P \propto \Sigma^{\Gamma}$ with adiabatic index $\Gamma=5/3$, 
to perform the modal analysis
on the analytical disk profiles studied by
\citet{1999ApJ...513..805L}.

A locally isothermal equations of state is used
to calculate unstable modes from
the numerical simulations
described in Section~\ref{sec:codes}.
The vertically integrated pressure is given by
\begin{equation}
P=c_{\mathrm{s}}^2 \Sigma,
\label{eq:isot}
\end{equation}
where $c_{\mathrm{s}}$ is the sound speed.
The temperature can be calculated from
the ideal gas equation of state
\begin{equation}
   P = \frac{\mathcal{R} \Sigma T}{\mu},
\end{equation}
where $\mathcal{R}$ is the gas constant and $\mu$ is the mean atomic weight.
The initial unperturbed density is uniform in our simulations
and the temperature has a fixed profile $\propto r^{-1}$ for the isothermal
calculations, where $r$ is the distance from the rotation axis.
The sound speed profile of the disk is that of a standard
slightly flaring solar nebula with constant aspect ratio
\begin{equation}
   c_{\mathrm{s}} = 0.05 \sqrt{\frac{G M_*}{r}}
\end{equation}
where G is the gravitational constant and $M_*$ the
mass of the central star.

The linearized equations can be
reduced to a second order differential equation
for the enthalpy of the fluid, $\eta = \delta P/\Sigma$,
in the general case when the pressure is a function
of both density and temperature
\citep{1999ApJ...513..805L,2000ApJ...533.1023L,
2004ASPC..321..333D}.
\begin{equation}
  \eta^{\prime\prime} + B(r) \eta^\prime + C(r) \eta = 0
  \label{eq:ODE}
\end{equation}
where the coefficients $B(r)$ and $C(r)$ depend on the radial distance,
\begin{eqnarray}
  B(r) & = &
    \frac{1}{r} + \frac{{\cal F}'}{\cal F} -
    \frac{\Omega'}{\Omega} \\
  C(r) & = & -k_\phi^2
    - \frac{\kappa^2 - \Delta \omega^2}{c_{\mathrm{s}}^2}
    - 2 k_\phi \frac{\Omega}{\Delta \omega}
    \frac{{\cal F}'}{\cal F} \\
    & &-\frac{1-L_s'}{L_s^2}  
    -\frac{B(r)+4 k_\phi \Omega / \Delta \omega}{L_s} 
    -\frac{k_\phi^2 c_{\mathrm{s}}^2 / \Delta \omega^2 - 1}{L_s L_p}
\end{eqnarray}
and
\begin{eqnarray}
   {\cal F}(r) &=& \frac{\Sigma \Omega}{\kappa^2 - \Delta \omega^2 -c_{\mathrm{s}}^2 /( L_sL_p)}\\
  k_\phi & = & \frac{m}{r}\\
  \kappa^2 & = & \frac{1}{r^3}\frac{\ud l^2}{\ud r}\\
  \Delta \omega(r) &=& \omega_r+i\gamma-m\Omega(r)
\end{eqnarray}
where $\kappa$ is the epicyclic frequency, $l=\Omega r^2$ is the
angular momentum per unit mass,
$c_{\mathrm{s}}$ the adiabatic sound speed 
and $L_s$, $L_p$ are the equilibrium length
scales of variation of entropy
and pressure changes, given by:
\begin{eqnarray}
   L_s^{-1} & = & \frac{1}{\Gamma} \frac{\ud \mathrm{ln} (P/\Sigma^\Gamma)}{\ud r} \\
   L_p^{-1} & = & \frac{1}{\Gamma} \frac{\ud \mathrm{ln} (P)}{\ud r}.
\end{eqnarray}
$L_s$ and $L_p$ are calculated numerically
from the averaged profiles obtained in 
the simulations using Equation~\ref{eq:isot}.

The growth of the unstable modes can form vortices
or Rossby waves
in the nonlinear regime \citep{2001ApJ...551..874L}.
The Rayleigh criterion states that the disk will be
stable to axisymmetric perturbations when the specific angular
momentum increases with radial distance.
For a Keplerian disk the epicyclic frequency is
always positive, $\kappa^2=\Omega_{\mathrm{K}}^2$,
where the Keplerian angular frequency is given by $\Omega_{\mathrm{K}} = (GM_*/r^3)^{1/2}$,
and therefore axisymmetric waves will be stable.
However, when the pressure effects are taken into account it
is possible to have axisymmetric instabilities for a sufficiently
large pressure gradient according to the Solberg-H\o iland criterion,
\begin{equation}
\kappa^2(r)+N_r^2(r) \geq 0
\label{eq:s-h}
\end{equation}
where
\begin{equation}
N_r^2(r) = - \frac{1}{\Gamma \Sigma}\frac{\ud P}{\ud r}
\frac{\ud }{\ud r} \mathrm{ln} \left(\frac{P}{\Sigma^{\Gamma}}\right)
\end{equation}
is the square of the Brunt-V\"ais\"al\"a frequency in
the radial direction.

\begin{figure}
\resizebox{\hsize}{!}{\includegraphics{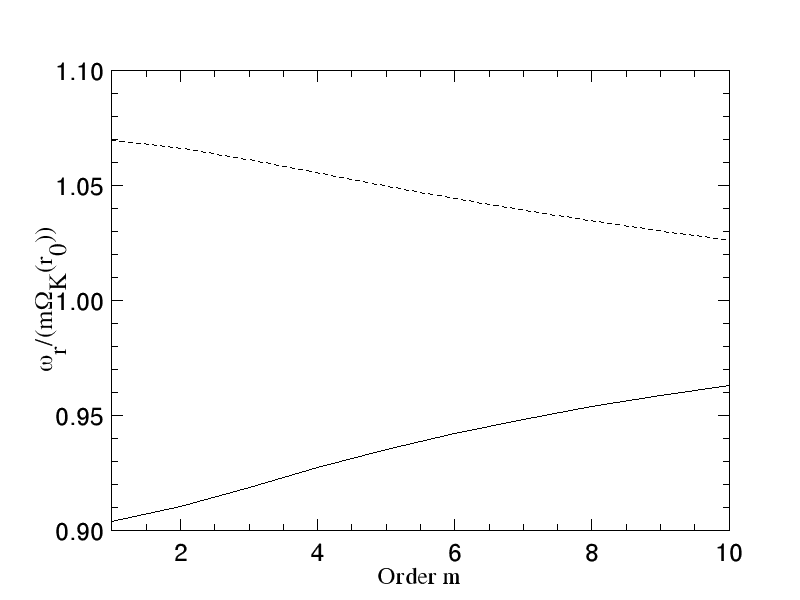}}
  \caption{Real frequency of the most unstable modes for  
  a gap opened by a Jupiter-mass planet after 10 periods
  as a function of the azimuthal mode number.
  The solid line shows the mode frequency at the outer edge of the gap 
  and the dashed line is the mode frequency at the inner edge 
  divided by $m \Omega_{\mathrm{K}}(r_0)$ where $\Omega_{\mathrm{K}}(r_0)$ is the
  Keplerian angular frequency at the planet position.
           }
  \label{fig:wreal}
\end{figure}

\begin{figure}
  \resizebox{\hsize}{!}{\includegraphics{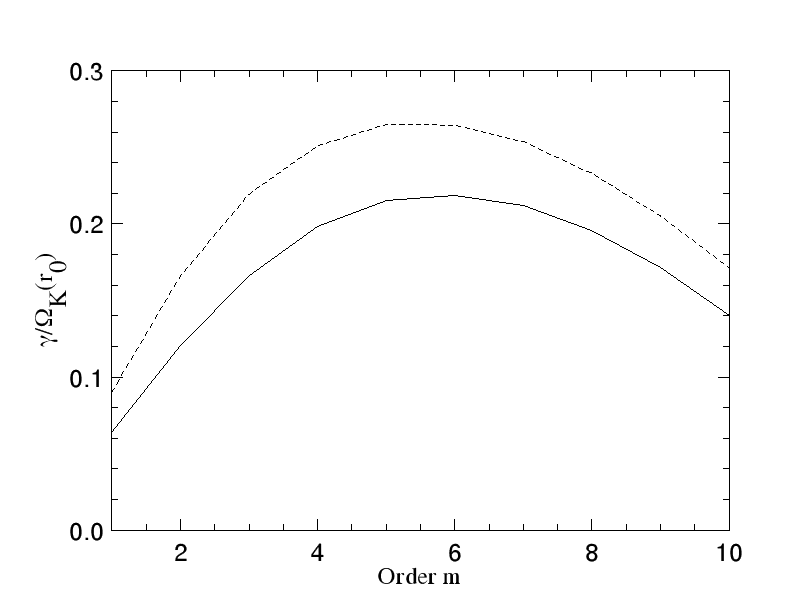}}
  \caption{Growth rate for a gap opened by a Jupiter-mass planet
  after 10 orbits against the azimuthal mode number.
  The solid line shows the growth rate at the outer edge of the gap 
  and the dashed line is the growth rate at the inner edge.
  The growth rate of the instability peaks at mode numbers 5--6.
  }
  \label{fig:wimag}
\end{figure}

The eigenproblem for the perturbed enthalpy
is solved using two semi-analytical methods. One
way of finding the complex eigenfrequency
is the shooting method, where the integration
proceeds from the disk boundaries to an intermediate fitting point, where
continuity of the eigenfunction and its first derivative
is required.
Our implementation uses a leapfrog method to integrate the equation from the
boundaries to the fitting point. The values of the entalphy and its
derivative at the starting points are specified 
based on several prescriptions using outgoing spiral waves
\citep{2000ApJ...533.1023L}
and vanishing eigenfuntion at the boundaries.
We find that the obtained growth rates do not depend
sensitively on
the choice of boundary conditions.
Several root finding algorithms in the complex plane
can be used to find the unstable modes.
The winding number theorem uses closed-path integrals
\citep[e.g.][]{1989ASAJ...85.1014K}
to find the number of roots inside the contour.

The winding number theorem states that for a complex analytic function
$f(\omega)$ defined inside a contour $C$
\begin{equation}
2\pi i(N-P) = \oint \frac{f'(\omega)}{f(\omega)} \, \ud \omega
\end{equation}
where $N$ is the number of roots and $P$ is the number of poles inside
the contour $C$, and
the integral is defined in counterclockwise sense.
We look for contours that comprise a single root and are not located
close to solutions or singularities on the real axis.
There are singularities in equation~\ref{eq:ODE} at the
corotation resonance $\Delta \omega = 0$ and when a modified
form of the Lindblad resonance for non-barotropic flow is satisfied,
$\kappa^2 - \Delta\omega^2 - c_{\mathrm{s}}^2/(L_s L_p) = 0$.
It is important to avoid branch points close to the contours
since the winding number method would not give the right
number of roots.
In the case of analytical density profiles given by hyperbolic
trigonometric functions the branch points do not affect the contours
used to find solutions with positive growth rate.

The solutions are then calculated by
\begin{equation}
\sum_{i=1}^{N} \zeta_i - \sum_{j=1}^{M} \xi_j = 
   \frac{1}{2\pi i}\oint \frac{\omega f'(\omega)}{f(\omega)} \, \ud \omega
\end{equation}
where $\zeta_i$ are the zeros and $\xi_j$ the poles of the complex function
$f(\omega)$.
The mode frequency can be determined in a contour with a single root.
A multidimensional Newton-Raphson method 
\citep{1992nrfa.book.....P}
is then employed to locate the roots with arbitrary
accuracy.

Another approach to solve equation~\ref{eq:ODE}
involves discretizing the equation
on a finite
grid and use appropriate boundary conditions to reduce the problem to finding
numerically the roots of the determinant of a complex tridiagonal matrix
\citep[e.g.,][]{1998ApJ...504..945L,2000ApJ...533.1023L}.
We solved the determinant using the previous root finding algorithm
and obtain the radial profile for the eigenfunction $\eta(r)$
and the perturbed variables.

We checked these two methods on the
axisymmetric analytical step jump profiles in surface density studied by
\citet{2000ApJ...533.1023L}.
We considered azimuthal mode numbers from $m=1$ to 10 and calculated the growth
rates of the unstable modes and the corresponding eigenfunctions. For
analytical density profiles with various shapes
both our methods agree with the results of
\citet{2000ApJ...533.1023L}.
We tested the dependence of the solution on the shape
of the pressure profile and aspect ratio of the disk.
The eigenfunction for density profiles
with a locally isothermal equation of state is
obtained by solving the discretized equation~\ref{eq:ODE}.

In Fig.~\ref{fig:wreal}, we show the real part of the mode frequency 
as a function of the azimuthal number for the
averaged density profiles of a \nirvana{} simulation (see Section~\ref{sec:codes})
at 
$n_\mathrm{r} \times n_{\phi}=256\times768$ resolution
after 10 orbits.
Fig.~\ref{fig:wimag} shows the growth rate as a function
of the mode number for the
same averaged density profile,
after 10 orbital periods when the threshold for the excitation
of the instability is reached.
The density slope in the simulations is described
using two parameters. The depth of the gap is calculated
with respect to the unperturbed density in the inner and
outer disk.
The length scale over which the density varies
is estimated to be the difference between the
local maximum and minimum at both inner and
outer disks.
An analytical jump function is fitted to the averaged profiles. 

The local maximum at the planet position in the
averaged profiles is not expected
to change the growth rates significantly.
In addition, most of the gap is clean,
apart from the material close to the planet position and at
Lagrangian points $L_4$, and $L_5$.
In the simulations with an initial gap
the averaged profiles do not have local maxima inside
the gap after 100 orbits.

The larger growth rates
in Fig.~\ref{fig:wimag} correspond to the more
unstable modes of equation~\ref{eq:ODE}
that will dominate the solution.
The inner edge of the gap opened by a planet,
shown by the dashed line, has higher growth rate than that of the
outer edge, represented by the solid line, at a given time.
This difference can be due to the fact that the inner
boundary is closer to the location of the instability
in the inner disk than it is the outer boundary to the corresponding 
instability outside the planet's radius.
The real frequency of the modes correspond to a radial location
just outside the corotating region where
vortices are formed in the simulations after about 10 orbits.
The number of vortices in the numerical results is consistent
with the growth rates peaking at $m = 4$--$6$.

\begin{figure}
  \resizebox{\hsize}{!}{\includegraphics{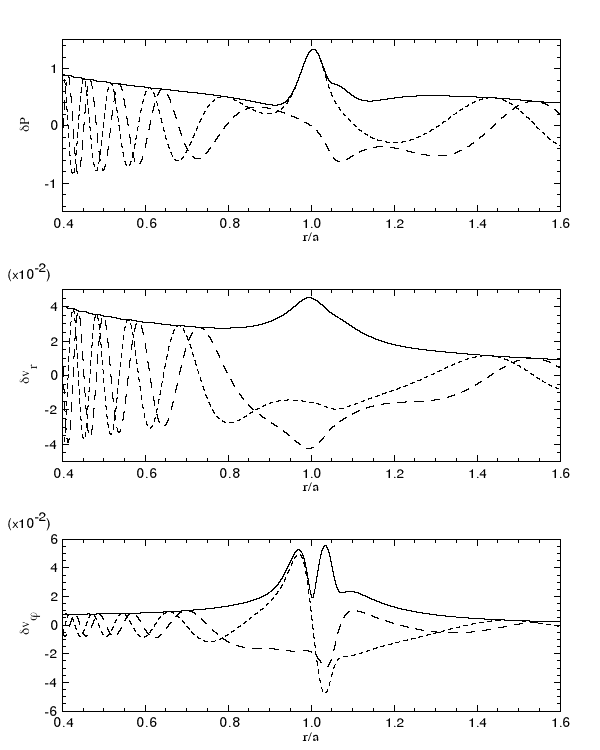}}
  \caption{Radial profile of the eigenfunctions for the outer
  edge of the gap at $t = 10$ periods and mode number $m = 5$.
  From top to bottom the pressure perturbation and radial
  and azimuthal perturbed velocity components are shown.
  The dotted and dashed lines are the real and imaginary part
  of the eigenfunctions. The amplitude is shown by the solid line
  which peaks at the position of the edge for the eigenfunction
  of the perturbed pressure.
  }
  \label{fig:eigen_rsj}
\end{figure}

\begin{figure}
  \resizebox{\hsize}{!}{\includegraphics{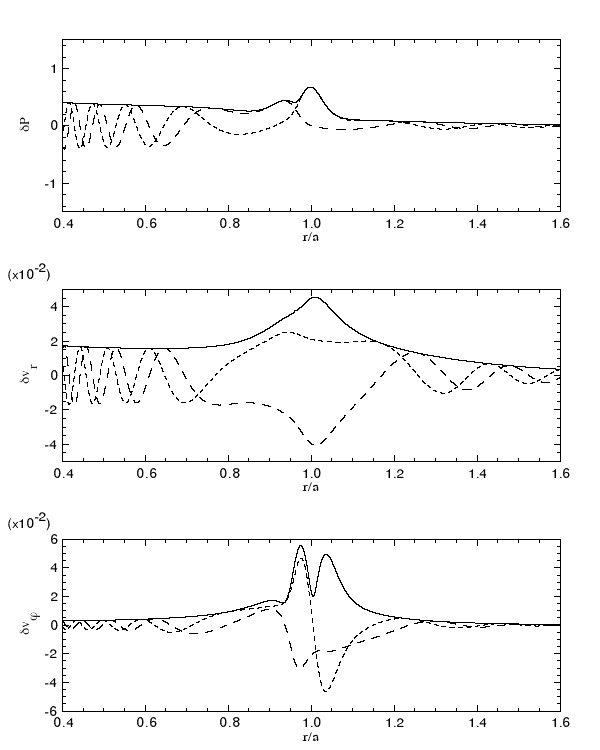}}
  \caption{Radial profile of the eigenfunctions for the inner
  edge of the gap at $t = 10$ periods and mode number $m = 5$.
  From top to bottom the pressure perturbation and radial
  and azimuthal perturbed velocity components are shown.
  The dotted and dashed lines are the real and imaginary part
  of the eigenfunctions.
  }
  \label{fig:eigen_lsj}
\end{figure}

In Fig.~\ref{fig:eigen_rsj} the real and imaginary
parts of the radial eigenfunction of the perturbed
variables at the outer edge
of the gap are plotted for azimuthal mode $m=5$.
The eigenfunction corresponds to a time when the gap
becomes deep enough to generate modes with positive
growth rate. The middle and bottom panel show the
eigenfunctions of the perturbations of the velocity components.
The radial eigenfunctions at the inner edge
of the gap after 10 orbits
for azimuthal mode $m=5$
are plotted in
Fig.~\ref{fig:eigen_lsj}.

\section{Numerical codes}
\label{sec:codes}

We performed 2-dimensional hydrodynamical simulations using two independent
grid-based codes, implemented in cylindrical and Cartesian coordinates.
The simulations were run on a uniform grid for 100 orbital periods
Different boundary conditions were tested to avoid reflection of
waves at the boundaries.

\subsection{Initial setup}
\label{sec:setup}

The computations were performed in the
radial domain $0.4 a \le r \le 2.5 a$
where $a$ is the planet semi-major axis.
The disk is assumed to be geometrically thin
and the fluid equations are solved
using vertically-integrated variables:
\begin{equation}
\Sigma(r,\phi) = \int_{-H}^{H} \rho(r,\phi,z) \, \ud z,
\qquad P(r,\phi)= \int_{-H}^{H} p(r,\phi,z) \, \ud z$$
\end{equation}
where $H$ is the vertical scale height.

The planet's gravitational potential was given by the formula
\begin{equation}
\phi = \frac {-\mu}{\sqrt{{s}^{2} + \epsilon^{2}}}
\end{equation}
where $\epsilon$ is the gravitational softening
and $s$ is the distance from the planet.
The softening is defined as
$\epsilon = 0.6 H_\mathrm{p}$,
with $H_\mathrm{p}$ the disc scale height at the planet location.
This softening was introduced to reproduce the
torque cut-off due to 3-dimensional effects
at distance from the planet larger than $H_\mathrm{p}$.
The softening length is
comparable in size to the characteristic size of the
Roche lobe (along the planet's orbit) and thus it is likely to
affect the flow dynamics in the corotation region but
it is unlikely to strongly affect the dynamics in the regions
where we observe vortex formation.
Initially, the gas rotates with Keplerian
angular frequency around the central star.
The mass ratio had the values
$\mu = M_\mathrm{p}/M_\mathrm{*} = 10^{-3}$ and $10^{-4}$,
corresponding to Jupiter and Neptune masses when
the stellar mass is one solar mass.
Self-gravity of the
disk was not included in the simulations.
The planet was kept in a coplanar, circular orbit
at $a=1$.
We did not consider the accretion
of disk material onto the protoplanet.

We assume that the disk radiates efficiently the thermal
energy generated from tidal dissipation and viscous heating.
In the absence of a radiation mechanism the gas would 
heat up and the disk could become geometrically thick.
In our models, we use a locally isothermal equations of state,
\citep{1985prpl.conf..981L}
with constant aspect ratio $H/r = 0.05$.

The vortensity or potential vorticity is calculated
in the corotating frame:
\begin{equation}
\zeta = (\nabla \times {\mvect{v}}+2\Omega_{\mathrm{p}})/\Sigma
\end{equation}
where $\Omega_{\mathrm{p}}$ is the orbital frequency of the planet.

We perform our simulations in the inviscid limit.
In addition, a few cases with
physical viscosity $\nu=10^{-6}$ and $10^{-5}$ 
$a^2 \Omega_{\mathrm{p}}$
are considered in order to estimate a threshold 
for the formation of vortices.
Artificial viscosity is not needed
to smooth the shocks in our codes, which is
important to study the formation of vortices.
The codes are also able to resolve the large density contrast
between the corotating region and the rest
of the disk.
Diffusion into the gap opened by the planet
may result from numerical viscosity or shocks.
The numerical viscosity
in our codes 
is estimated in Appendix~\ref{ap:viscosity}
for inviscid runs without tidal perturbations.
These tests confirm that numerical diffusion
is not affecting our results.

The initial density was uniform and the gravity
of the planet was introduced gradually
with the formula
\begin{equation}
M(t) = M_\mathrm{p} \sin^2 \left( \frac{\pi t}{10 P_\mathrm{p}} \right)
\end{equation}
We have used timescales between 5--10 orbital periods
to introduce the gravity from the planet. 
Although the time when the instability appears
depend on how the gravity is started, there is
agreement between the modal analysis and
numerical simulations for timescales of 5 and 10 orbits.
The results presented
in Section~\ref{sec:results} use 
a switch-on time of the gravity of 5 orbits.

In some of the calculations we use an initial gap profile
derived under the WKB approximation \citep[e.g.,][]{2006ApJ...641..526L}
\begin{equation}
\Sigma_0(r)= %
\exp{\left[- \frac{f}{9} \frac{q^2 a^2 \Omega_{\mathrm{p}} }{\nu_\mathrm{p}}\left(\frac{a}{\Delta_{\mathrm{p}}}\right)^3\right]},
\label{eq:init_gap}
\end{equation}
where we use $f/\nu_\mathrm{p}$ between 
$1.7\times10^{4}$ and $1.8\times10^{4}$
and $\Delta_\mathrm{p}$ is the maximum of 
$H$ and $|r-a|$.
This gap profile allows us to check whether a rapid formation of the gap
is necessary for the presence of instabilities.

The Cartesian implementation of \flash{}
was run on a uniform non-rotating grid at resolution
$n_\mathrm{x} \times n_\mathrm{y} =320\times320$,
and $n_\mathrm{x} \times n_\mathrm{y} =640\times640$.
The computational domain was
$-2.6a \le x \le 2.6a$
and $-2.6a \le y \le 2.6a$.

For computational convenience
the unit of time used in the simulations was the orbital period
at the planet location,
$P_\mathrm{p} = 2 \pi [a^3/G(M_\mathrm{*}+M_\mathrm{p})]^{1/2}=2 \pi$,
where $G (M_\mathrm{*}+M_\mathrm{p})=1$ and $a=1$.
Therefore, the angular frequency of the planet was $\Omega_\mathrm{p}=1$ in our units.

We used solid boundaries with wave damping zones next
to the boundaries to avoid wave reflection
that can create artificial resonances.
The damping regions were
implemented
in the \nirvana{} simulations
at 
$0.4a \le r \le 0.5a$ and $2.1a \le r \le 2.5a$,
by solving the following
equation after each timestep:
\begin{equation}
\frac {\ud x}{\ud t} = - \frac {x-x_\mathrm{0}} {\tau} R(r)
\end{equation}
where $x$ represents the surface density and velocity components,
$\tau$ is the orbital period at the corresponding boundary, and
$R(r)$ is a parabolic function which is one at the domain boundary
and zero at the interior boundary of the wave killing regions.
\nirvana{} simulations were also run using alternative boundary
conditions to check that the gap profiles and vorticity in the corotating
region do not depend strongly on boundary conditions.

\subsection{\nirvana}

Our \nirvana{} implementation is based on the
original version of the code by \citet{1997CPC..101..54}.
The Navier-Stokes equations are solved using
a directional operator splitting upwind scheme
which is second-order accurate in space and semi-second-order in
time.

\nirvana{} uses a staggered grid where scalar quantities are stored
at the center of grid cells and vectors are stored at
the cell boundaries.
The frame is centered on the center of mass of
the star-planet system and rotates with the
same angular frequency as the planet orbits
the central star.
The code was run with both non-reflective boundary conditions
described above \citep[see also][]{1996MNRAS.282.1107G}
and wave-killing zones close to the boundaries
\citep{2006MNRAS.370..529D}.
These two types of boundary conditions
provide consistent results since in both cases
wave reflection in the boundaries is reduced.
A Courant number of 0.5 was used in the simulations.

Two grid resolutions were considered in which the number of cells
in the radial and azimuthal directions were
$n_\mathrm{r} \times n_{\phi}=256\times768$
and $512\times1536$, respectively,
with uniform spacing
in both dimensions.
Therefore, the cells around the planet position
were approximately square.

\subsection{\flash}

The  \flash{} code \citep{2000ApJS..131..273F} is a fully parallel AMR implementation
of the PPM algorithm
in its original Eulerian form\footnote{\flash{} is available at \url{http://www.flash.uchicago.edu/}}
\citep{1984JCP...54..115,1984JCP...54..174}.
\flash{} has been extensively tested in various
compressible flow problems with astrophysical applications
\citep{2002ApJS..143..201C,2005Ap&SS.298..341W}.

Our implementations of the \flash{} code
used both polar coordinates and the original Cartesian formulation of \flash{}.
The polar version of \flash{} was run in the corotating
coordinate system while the Cartesian implementation
was run in the inertial frame.
Our implementations were based on release 2.5 of \flash{}
with customized modules for
the equation of state and gravity forces,
explicitly ensuring the conservative transport of angular momentum
in the angular sweep.
This is particularly important when large density
gradients are present in the disk.
The Coriolis forces were treated conservatively as described
by \citet{1998A&A...338L..37K}.
The isothermal Riemann solver was ported from the AMRA code
\citep{2001CPC..138..101}.
Courant numbers of 0.7 and 0.8 were used in the simulations.

The Cartesian grid was
centered in the center of mass of the system
and the orbits of the planet and the central star were
integrated using a simple Runge-Kutta method.
The grid cells were sized to
give the same radial resolution in
both implementations, although since the grid went to $r=0$
in the Cartesian simulations, the
grid size had to be larger to achieve the same
resolution.  
The Cartesian code was run at
resolutions $n_{\textrm{x}} \times n_{\textrm{y}} =320\times320$ and $640\times640$.
To improve the angular resolution close to
the central star, an additional level of refinement was used
in the inner disk in the Cartesian implementation.
Different timesteps were used for each level of refinement
to speed up the simulation.

The damping condition described in
\citet{2006MNRAS.370..529D}
was applied 
in the ring $2.1a \le r \le 2.5a$ 
close to the outer boundary
but not in
the inner disk
for the Cartesian \flash{} implementation.
A free outflowing boundary was used
at the outer boundaries $x = \pm 2.6a$ and $y = \pm 2.6a$.
and there was free
gas flow inside $0.4a$.

The cylindrical implementation used a frame
centered on the star
with resolutions
$n_r \times n_{\phi}=256\times768$
and $512\times1536$
and wave-damping zones close to the
inner and outer boundaries.
The Coriolis force, centrifugal force,
and indirect terms due to the fact that the center of the frame
is displaced from the center of mass
of the system
are included in the equation of motion.

\section{Results}
\label{sec:results}

We carried out modal growth analysis on the density profiles of
protoplanetary disks
perturbed
by an embedded giant planet
ranging from a Neptune to a Jupiter mass.
The time resolved linear analysis was performed on the
azimuthally averaged 
density profiles obtained from calculations, 
using a locally isothermal equation of state.
In the following description, we will show the results
of inviscid runs at different resolutions with an
embedded Jupiter-mass protoplanet using different
boundary conditions. 
Results from runs with physical viscosity will also be presented.

\begin{figure}
  \resizebox{\hsize}{!}{\includegraphics{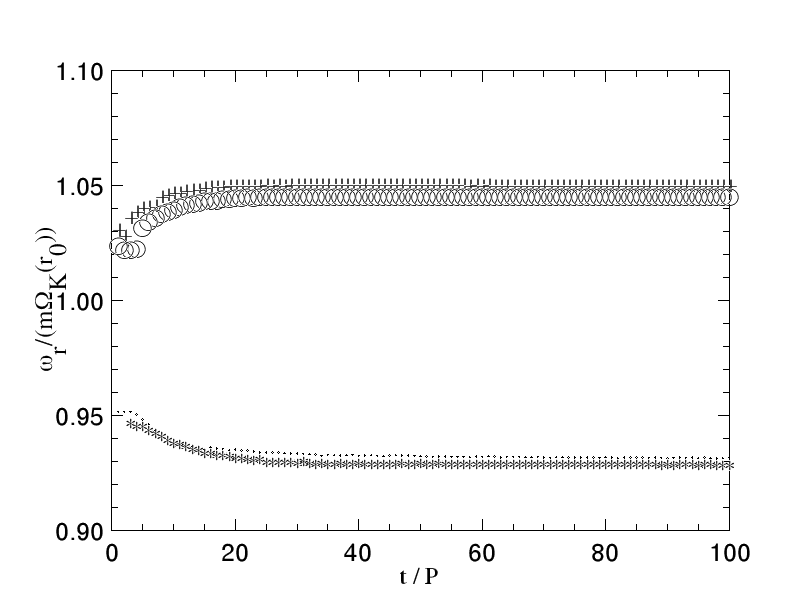}}
  \resizebox{\hsize}{!}{\includegraphics{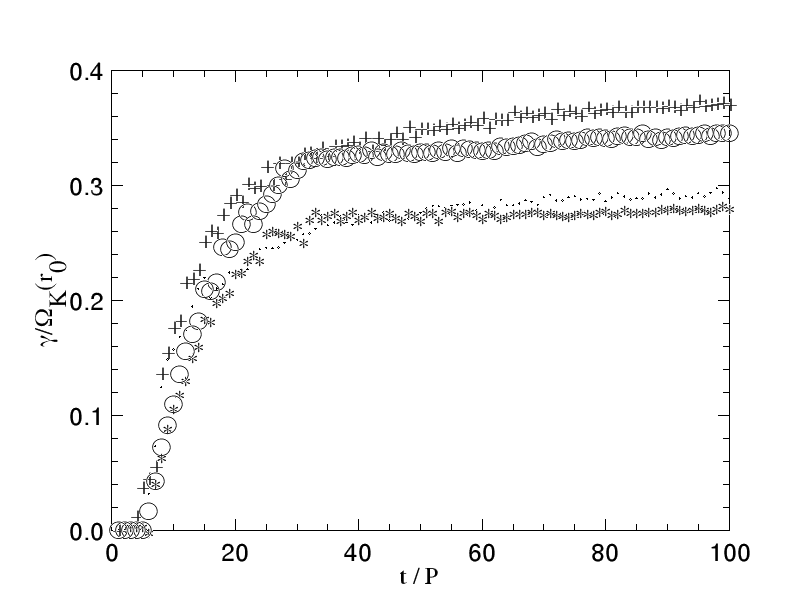}}
  \caption{Real frequency and growth rates of the unstable modes
  with azimuthal number $m=5$,
  as a function of time, for \nirvana{} and polar \flash{} simulations
  with resolution $n_r \times n_{\phi}=256\times768$.
  In the top panel, the crosses represent the mode frequencies
  at the outer edge of the gap
  and the dots are the mode frequencies at the inner edge for \nirvana{}.
  The circles are the mode frequencies at the outer edge of the gap
  and the stars are the mode frequencies at the inner edge for polar \flash{}.
  The frequencies are
  divided by $m \Omega_{\mathrm{K}}(r_0)$, where $\Omega_{\mathrm{K}}$ is
  the Keplerian angular frequency
  and $r_0$ is the radius at the edge of the gap.
  The crosses in the bottom panel are the growth rates 
  at the outer gap edge
  and the dots are the growth rates at the inner edge for a \nirvana{} calculation.
  The circles are the growth rates at the outer edge of the gap
  and the stars are the growth rates at the inner edge for a polar \flash{}
  calculation.
  Growth rates are divided by the Keplerian frequency
  at the edge of the gap.
  }
  \label{fig:w_nirv}
\end{figure}

The real frequency and growth rate
of the most unstable modes, as a function of
time, are plotted in Fig.~\ref{fig:w_nirv}
for \nirvana\ and \flash\ polar simulations
with resolution
$n_r \times n_{\phi}=256\times768$ and
wave-damping boundary conditions.
The threshold for the appearance of the instability
occurs after about 5 periods when the gap is sufficiently
deep and the growth rate
becomes positive.
The edge of the gap in the simulations
becomes non-axisymmetric at this time and
depressions
in vortensity appear along the gap with $m=5$--$6$ symmetry
which coincide with the azimuthal number of the
most unstable modes shown in
Fig.~\ref{fig:wimag}.
Small vortices are observed in
the inner and outer disk shortly afterwards
with the same angular distribution.

In the bottom panel of Fig.~\ref{fig:w_nirv},
the growth rate from the inner disk edge,
represented by crosses and circles
for the different models,
is larger than the growth rate at the outer disk,
represented by dots and stars.
This difference may be artificial because the inner
boundary of the grid is closer to the planet location
that the outer boundary is.
The mode frequencies and growth rates for \nirvana\ 
and polar \flash\
agree within 10\%.
This is in agreement with the
vortex sizes obtained from the
Fourier analysis of the gravitational torques on
the planet.
In the end of the simulation, the growth rate of the instability
is a fraction of the angular velocity
at the edge of the gap.
The disk becomes unstable
to axisymmetric perturbations at time $\sim 40$ orbits
for the Jupiter simulations according
to the Rayleigh criterion (Eq.~\ref{eq:s-h}).
The RWI grows exponentially
in agreement with the linear analysis
during the first orbits
\citep{2001ApJ...551..874L}
and produces vortices
that can be sustained by interaction
with the spiral arms generated by the planet.

\begin{figure*}
  \includegraphics[width=17cm]{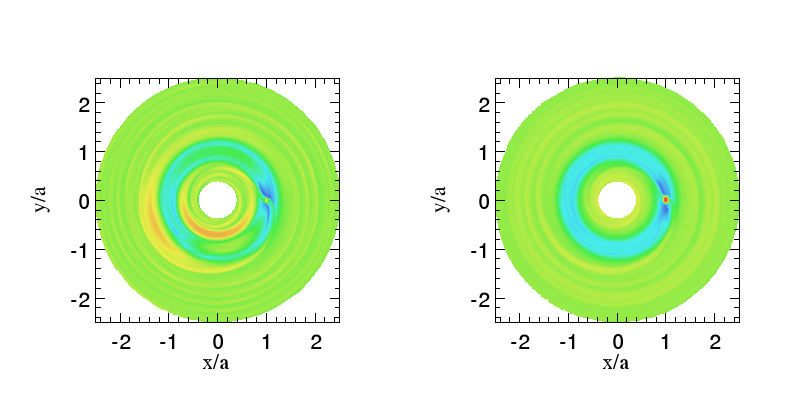}
  \caption{Surface density distribution after 100 orbital periods for \nirvana\
  on the left hand side and \flash\ on the right hand side
  using the same logarithmic color scale.
  Both models use
  the same wave damping condition in the outer disk between 2.1--2.5 $a$. 
  while \flash\ does not have a damping condition in the inner boundary.
  \nirvana\ has
  density enhancements close to the gap opened by the protoplanet.
  \flash\ has a smooth density distribution and a larger
  density peak at the
  planet position which is saturated in the image.}
  \label{fig:surfdens}
\end{figure*}

The linear analysis of Cartesian \flash{} averaged profiles
shows the presence of unstable modes in the inner
and outer disk with growth rates of
$\sim 0.2 \Omega_{\mathrm{K}}(r_0)$.
In some Cartesian \flash{} runs
there are mode solutions with
large growth rates that appear at late times,
when the gap is becoming deeper.
At the same time there are
indications that
some mild instability is happening in the disk. However, these
are not the fast-growing
modes found in the polar grid 
calculations with azimuthal number $m=4$--$6$.
We do not find stationary solutions with positive growth rates
that remain for a time of order of the growth timescale in the
Cartesian grid models. This is consistent with the fact that no
vortices appear in the
edges of the gap in these simulations after tens
of orbital periods, as shown in Fig.~\ref{fig:surfdens}.

\begin{figure*}
  \centering
  \includegraphics[width=17cm]{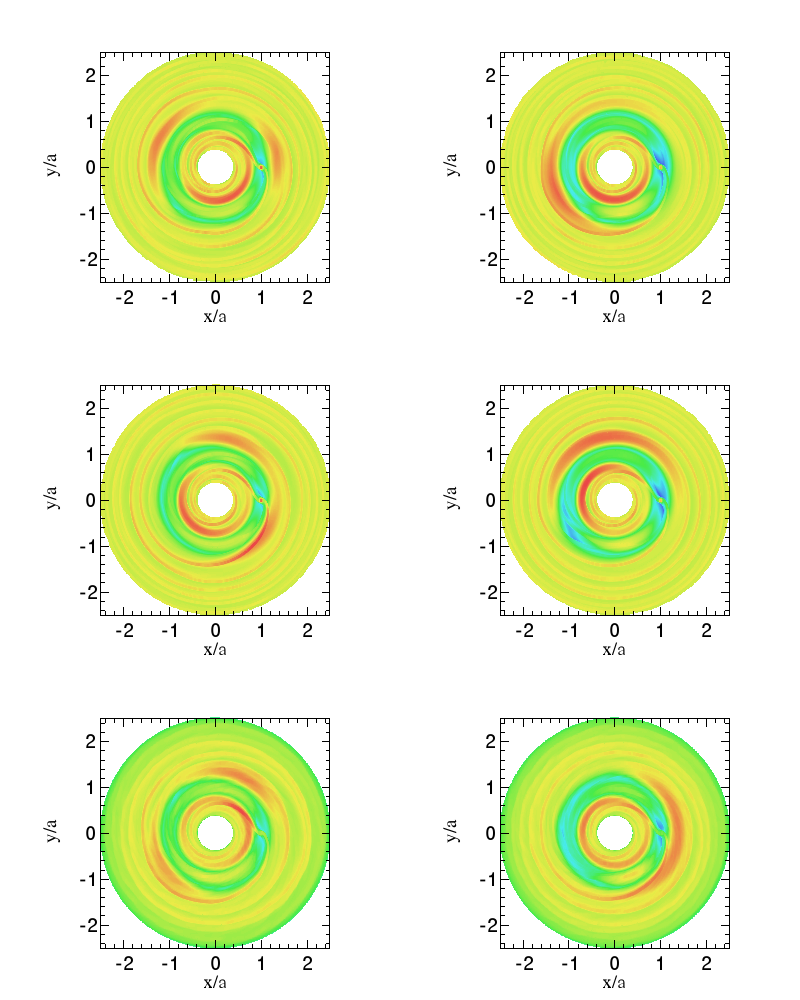}
  \caption{Surface density contours 
   for cylindrical simulations in logarithmic scale.
   From top to bottom, \nirvana\ simulation using a damping wave region
   as described in Section~\ref{sec:codes} \citep[see also][]{2006MNRAS.370..529D},
   \nirvana\ simulation using outgoing-wave boundary conditions
   defined by \citet{1996MNRAS.282.1107G}
   and polar \flash{}.
   The left panels show the density after 50 
   orbital periods
   and the right panels show the density after 100 orbits
   with the same color scale.
   The resolution is 
   $n_r \times n_{\phi} =256\times768$
   and the planet is located at
   $x \times y =(1,0)\,a$.
   There are two elongated vortices forming next to the outer gap edge
   at 50 orbits that move with different velocities. The vortices merge
   in both simulations and one large vortex is observed in the outer disk
   at 100 orbits.}
  \label{fig:nirvana}
\end{figure*}

Fig.~\ref{fig:nirvana}
shows the density distribution for Jupiter
inviscid simulations in logarithmic scale
using different boundary conditions for
\nirvana\ and \flash\ models.
The left panels show the density contours at $t = 50$ orbital
periods and the right panels after $t = 100$ orbits. There are two
vortices moving along the outer edge with different
phase velocities in all the simulations at $t = 50$ periods.
Those vortices will merge and form a large vortex at
a later time producing strong oscillations on the
torque exerted on the planet.

\begin{figure}
  \resizebox{\hsize}{!}{\includegraphics{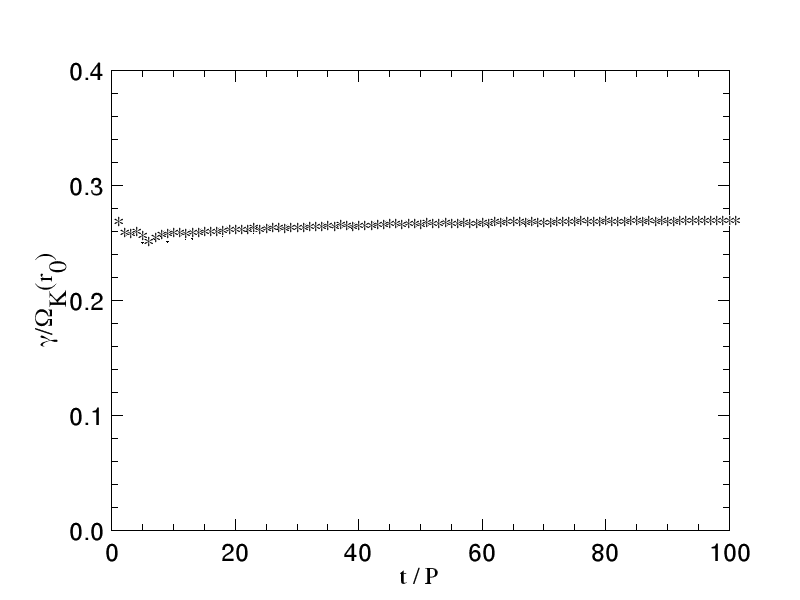}}
  \caption{Growth rates of the unstable modes
  with mode number $m=5$,
  as a function of time
  divided by the local Keplerian frequency.
  Dots represent a \nirvana{} calculation
  with resolution $n_r \times n_{\phi}=256\times768$
  and stars are obtained from a \flash{} calculation
  with resolution $n_r \times n_{\phi}=128\times384$.
  }
  \label{fig:nirvana_gap}
\end{figure}

In Fig.~\ref{fig:nirvana_gap}
we show the growth rates
with azimuthal number $m=5$
for the outer edge of the gap
using an initial density with a gap given by Equation~\ref{eq:init_gap}.
The gap shape tends to a steady state towards the
end of the simulation. The growth rates agree within
5\% (by the end of the simulation) with the results obtained using
an initial uniform density.

\begin{figure}
  \resizebox{\hsize}{!}{\includegraphics{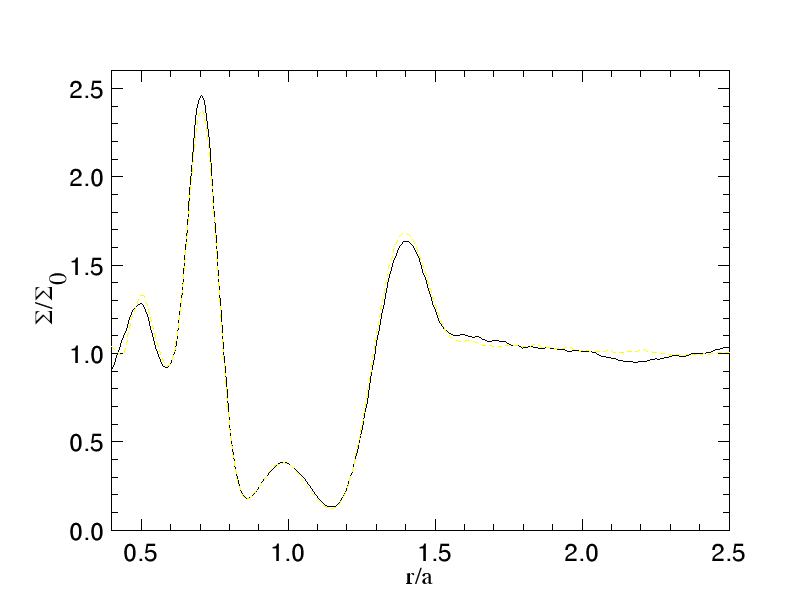}}
  \caption{Azimuthally averaged surface density profiles for the
   Jupiter simulations after 100 orbital periods.
   The solid line is the \nirvana{} simulation with resolution
   $n_r \times n_{\phi} =256\times768$
   and damping wave boundary conditions
   and the short dashed line
   is the \nirvana{} simulation with resolution $n_r \times n_{\phi} =256\times768$
   and outgoing-wave boundary conditions.
   The profiles are averaged over 5 orbits.
  }
  \label{fig:profile}
\end{figure}

Fig.~\ref{fig:profile} shows the profiles averaged over the azimuth
for the \nirvana{}
simulations using
various boundary conditions.
The slope of the gap is steeper at the inner edge
than at the outer edge for the cylindrical schemes.
The density peak in the inner disk is about
50\% of the peak at the outer gap edge.
The overall jump in density is larger at the inner edge despite the fact that
the gap is slightly deeper just outside the planet's orbit.
This produces a larger growth rate
inside the corotation region (see Fig.~\ref{fig:wimag}).
In the cylindrical \flash{} simulation,
the mass loss at the boundaries is greater 
but this does not affect the results of the modal calculation.
The Cartesian \flash{} results have a deeper gap and smaller
density peaks at the edge of the gap. However, the
growth rate of their most unstable modes is about 50\%
compared with that provided by the polar \flash\ simulations.

The size of the peaks in the power spectrum of the
torques on a Jupiter-mass planet,
from different regions in the disks,
are shown
in Tables~\ref{table:pds_jup} and~\ref{table:pds_jup_outer}.
The torques are calculated excluding the material
inside the Roche lobe of the planet, where the
resolution may not be good enough to resolve the
circumplanetary disk.
\nirvana\ calculations have a larger peak, which correspond to
the amplitude of the vortices moving along the edge of the gap.
This is consistent with larger vortices being
observed in the density distributions of \nirvana\ calculations
in Fig.~\ref{fig:nirvana}.
\flash\ calculations have PDS peaks which
are between one and two
orders of magnitude smaller.
The frequencies in the corotating frame
are close to the local Keplerian angular frequency
at the edges of the gap.
The difference in the peak amplitudes
agrees with the results from the
upwind and Godunov schemes
studied by \citet{2006MNRAS.370..529D}.
This correlation
suggests that vortices can be formed by the non-linear evolution of
Rossby waves in protoplanetary disks.

\begin{table}
   \caption{The frequency in the corotating frame
   and magnitude of the maxima of the PDS of the
   gravitational torques from the inner disk are shown for the Jupiter 
   simulations.
   The values are sorted by the
   magnitude of the PDS at the maximum. The magnitude of the PDS is related
   with the amplitude of the oscillations at a given frequency. Polar codes
   have larger maxima and the frequencies correspond roughly to the Keplerian
   frequencies at the gap edges.}
   \label{table:pds_jup}
   \centering
   \begin{tabular}{c c c}
      \hline\hline
      Code & Frequency & Amplitude \\
      \hline
      
NIRVANA &  0.568 &  $2.81\times10^{-5}$ \\
Flash polar&   0.531 &  $8.96\times10^{-8}$ \\
Flash Cart.&   1.611 &  $2.19\times10^{-7}$ \\

      \hline
   \end{tabular}
\end{table}

\begin{table}
   \caption{The frequency in the corotating frame
   and magnitude of the maxima of the PDS of the
   gravitational torques from the outer disk are shown for the Jupiter
   simulations.}
   \label{table:pds_jup_outer}
   \centering
   \begin{tabular}{c c c}
      \hline\hline
      Code & Frequency & Amplitude \\
      \hline
      NIRVANA  & 0.389 &    $1.11\times10^{-3}$ \\
Flash polar &   0.353 &   $4.82\times10^{-6}$ \\
Flash Cart &   0.997 &   $3.65\times10^{-6}$ \\

      \hline
   \end{tabular}
\end{table}

\begin{table}
   \caption{The frequency and magnitude of the maxima of the PDS
   of the torques from the inner disk for the Neptune case are
   shown sorted by the
   magnitude of the PDS at the maximum. \nirvana\ calculations have
   larger maxima with frequencies close to the Keplerian
   frequencies at the outer gap edge.}
   \label{table:pds_nep}
   \centering
   \begin{tabular}{c c c}
      \hline\hline
      Code & Frequency & Amplitude \\ 
      \hline
      
NIRVANA &  0.204 &  $4.67\times10^{-9}$ \\
Flash polar &   0.221 &   $3.57\times10^{-8}$ \\
Flash Cart.&   0.345 &  $3.99\times10^{-8}$ \\

      \hline
   \end{tabular}
\end{table}

\begin{table}
   \caption{The frequency and magnitude of the maxima of the PDS
   of the torques from the outer disk for the Neptune case are
   shown sorted by the
   magnitude of the PDS at the maximum.}
   \label{table:pds_nep_outer}
   \centering
   \begin{tabular}{c c c}
      \hline\hline
      Code & Frequency & Amplitude \\ 
      \hline
      NIRVANA  & 0.389 &    $1.11\times10^{-7}$ \\
Flash polar&   0.204 &  $5.27\times10^{-8}$ \\
Flash Cart &   3.997 &   $5.411\times10^{-7}$ \\

      \hline
   \end{tabular}
\end{table}

\begin{figure*}
 \centering
  \includegraphics[width=17cm]{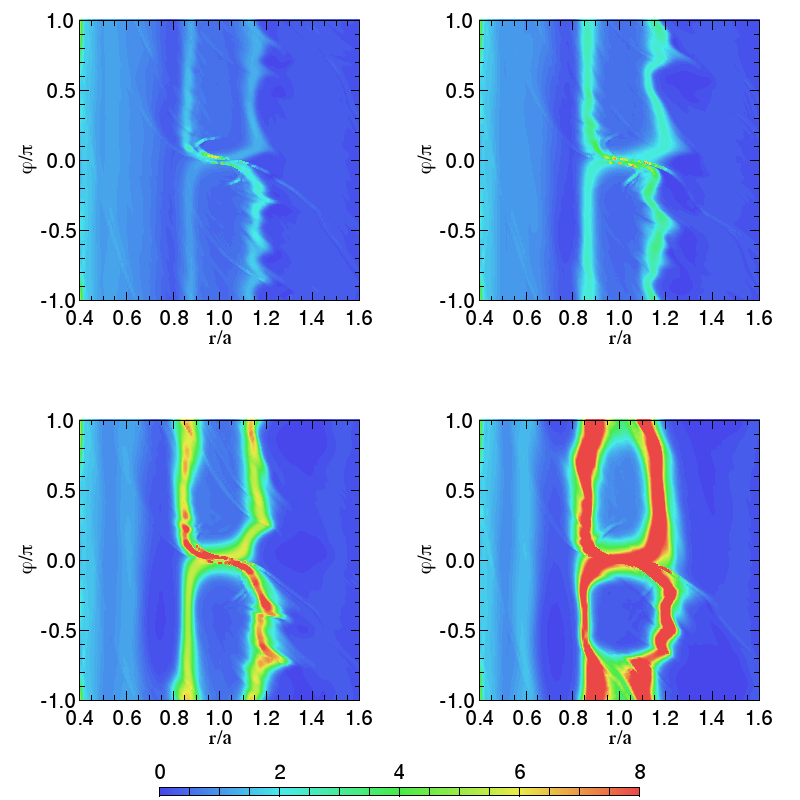}
  \caption{Vortensity in polar coordinates is shown at times
  $t = 10, 20, 50$ and $100$ orbital periods from left to right and top to bottom.
  The simulation has resolution $n_\mathrm{r} \times n_\phi = 512 \times 1536$.
  }
  \label{fig:vort_nirvana}
\end{figure*}

The evolution of the vortensity
in the \nirvana{} simulation with resolution
$n_\mathrm{r} \times n_\phi = 512 \times 1536$
is shown at several times in
Fig.~\ref{fig:vort_nirvana}.
The vortensity and Bernoulli constant are conserved for a barotropic
inviscid fluid in the absence of discontinuities in the flow.
In our case, the vortensity is roughly conserved along the
streamlines in regions outside the Hill radius of the planet.
Shock dissipation close to the protoplanet can lead
to vortensity generation.
As the gap is opened and strong trailing shocks are formed,
the vortensity
grows at the
edge of the gap and along the spiral arms.
The vortensity at 10 orbital periods shows small cavities outside
the peaks at the edge of the gap. These depressions
break in 4-5 differentiated vortices
when the growth rate of the RWI becomes positive.
The vortensity peak at the outer gap edge and close to the outer spiral arm
are corrugated while at the inner edge the vortensity
is more stable. As explained before, this is probably an artificial effect
due to the inner edge of the gap
being closer
to the inner boundary
than the outer gap edge is to the outer boundary.
The minima of vortensity rotating along the edge
correspond to vortices observed
in the density maps.
In the bottom right panel of Fig.~\ref{fig:vort_nirvana}
there is one single
vortensity depression
which is associated with a
vortex located at azimuth $\approx \pi$
after 100 orbits.
The vortensity
inside the corotating region is considerably perturbed
as the vortex moves along the gap.

\begin{figure}
  \resizebox{\hsize}{!}{\includegraphics{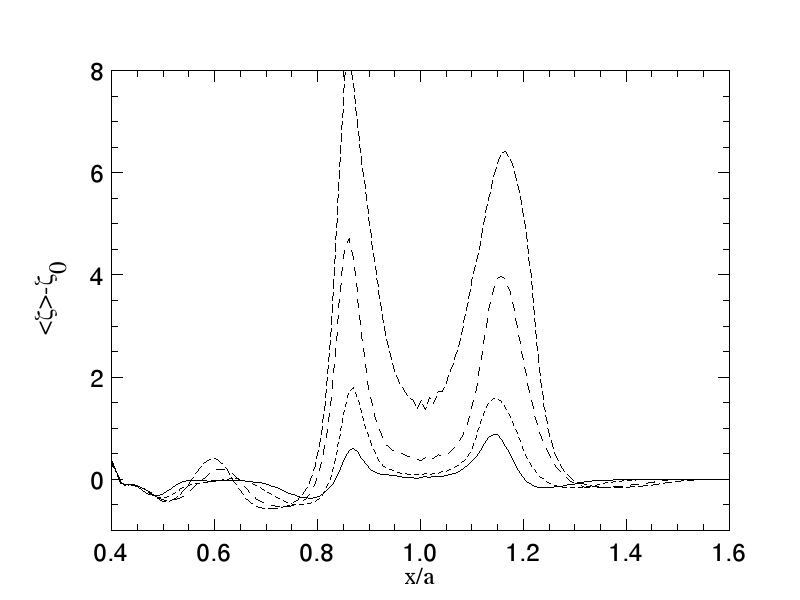}}
  \caption{Vortensity profiles averaged over azimuth at different times
  $t = 10, 20, 50$ and $100$ orbital periods for the \nirvana{} simulation
  at resolution $n_\mathrm{r} \times n_\phi = 256 \times 768$.}
  \label{fig:prof_vort_nirvana}
\end{figure}

\begin{figure}
  \resizebox{\hsize}{!}{\includegraphics{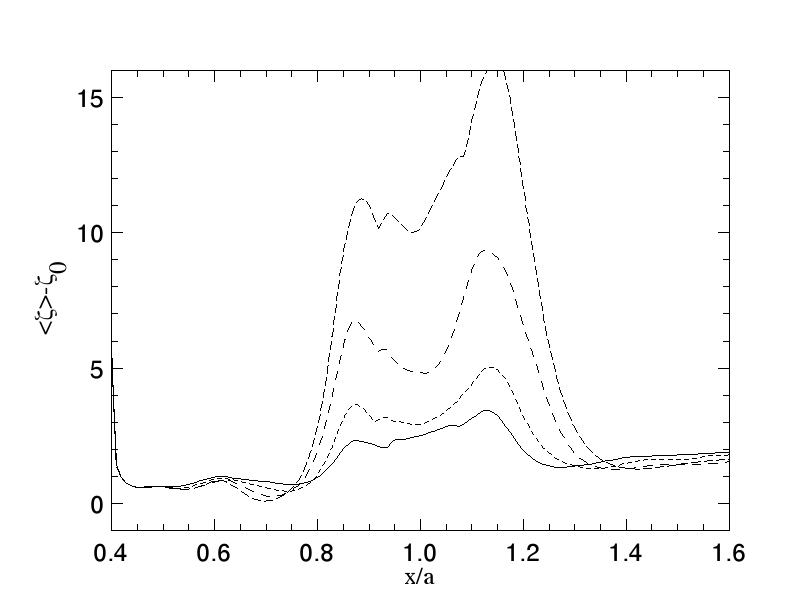}}
  \caption{Vortensity profiles averaged over azimuth at different times
  $t = 10, 20, 50$ and $100$ orbital periods for the FLASH simulation
  in Cartesian coordinates.}
  \label{fig:prof_vort_flash}
\end{figure}

In Figs.~\ref{fig:prof_vort_nirvana}
and~\ref{fig:prof_vort_flash}
the azimuthally averaged vortensity in the inertial frame is shown.
The initial vortensity profile for a disk with uniform density,
$\zeta_0 \propto r^{-1.5}$,
has been subtracted.
The Cartesian \flash\ code has a large vortensity excess 
in the corotating region where the gap is more depleted
than in the cylindrical codes.
The averaged vortensity in the outer disk is also greater
in our Cartesian \flash\ model.
\nirvana\ calculation shows vortensity peaks at the gap borders with
a larger spike in the inner disk.

\begin{figure*}
  \includegraphics[width=17cm]{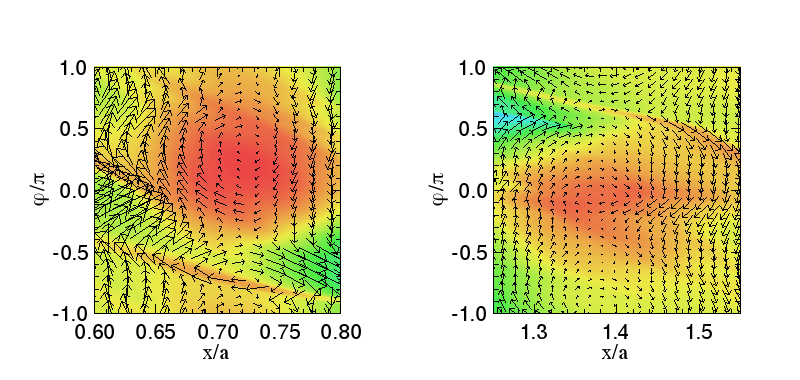}
  \caption{Velocity vectors in the corotating frame after 100 
  periods for the \nirvana{} simulation at resolution $n_\mathrm{r} \times n_\phi = 512 \times 1536$.
  The left panel shows the vortex at the inner gap edge
  and the right panel shows the vortex at the outer gap edge.}
  \label{fig:vortex2}
\end{figure*}

\begin{figure*}
  \includegraphics[width=17cm]{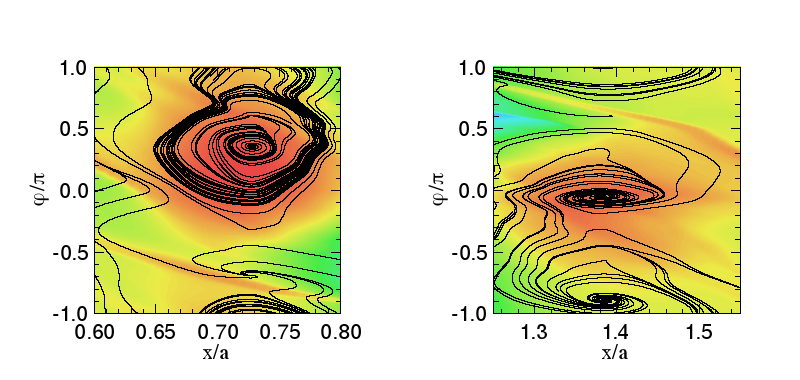}
  \caption{Streamlines in the corotating frame
   of the inner and outer vortices plotted in Fig.~\ref{fig:vortex2}.
  The radial extent of the vortices is about $0.15a$.
  Spiral arms created by the planet and weaker
  shocks associated with the vortices are observed.
  }
  \label{fig:vortex2_stream}
\end{figure*}

The velocity fields plotted over
the density contours in  logarithmic scale at $t = 100$ orbits
is shown in Fig.~\ref{fig:vortex2}. 
The velocity vectors are calculated in the corotating
frame of the local maximum of pressure that coincides
with the center of the vortex.
The rotation in baroclinic sense is very clearly
visible in the vortex close to the inner edge of the gap.
The vortex in the outer disk interacts with the
spiral wake created by the planet and is
perturbed at this particular time.
In Fig.~\ref{fig:vortex2_stream},
the streamlines are shown in the
frame corotating with the
vortex core.
They show baroclinic rotation that is perturbed
by interaction with the spiral wakes created
by the planet.

\subsection{Dependence on physical viscosity}

In this Section we compare the unstable modes
obtained from simulations including  Navier-Stokes viscosity
$\nu=10^{-6}$ and $10^{-5}$ (in code units, see Section~\ref{sec:codes}).

\begin{figure}
  \centering
  \resizebox{\hsize}{!}{\includegraphics{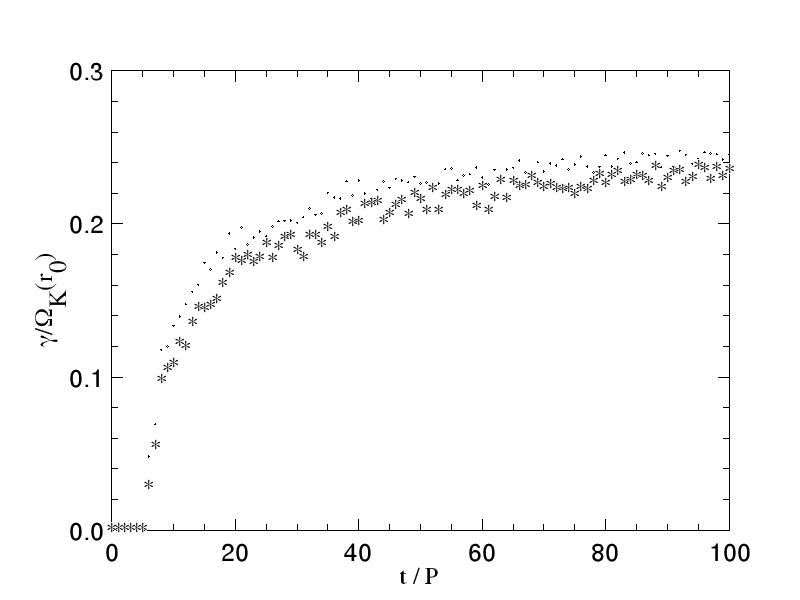}}
  \caption{
   Dependence on time of growth rates
   of modes with $m=5$
   for \nirvana{} simulations
   with resolution $n_\mathrm{r} \times n_\phi = 256 \times 768$ (dots)
   and \flash{} with resolution $n_\mathrm{r} \times n_\phi = 128 \times 384$ (stars).
   The Navier-Stokes viscosity is $\nu=10^{-6}$.
          }
  \label{fig:nirvana_visc}
\end{figure}

In Fig.~\ref{fig:nirvana_visc}
we show the growth rates in the outer disk of the
unstable modes with $m=5$
as a function of time
for $\nu=10^{-6}$.
The frequencies of unstable modes are
calculated using perturbations on the inviscid
Euler equations for simplicity.
The growth rates at the outer edge of the gap agree within 20\% with those obtained from inviscid
simulations (see Fig.~\ref{fig:w_nirv}).
However, notice that the linear analysis gives only an estimate
of the growth rates in the viscous case.

\section{Discussion}

We have studied vortex formation in
protoplanetary disks with an embedded giant planet,
with mass ratios  $10^{-4}$ and $10^{-3}$, using
numerical simulations and linear perturbation analysis.
The modal calculation is done following the
strategy of \citet{1999ApJ...513..805L}
for a locally isothermal equation of state
in a vertically-averaged disk.
Vortices are formed in the cylindrical \nirvana{} and \flash{}
2-dimensional simulations
in agreement with the linear analysis of non-axisymmetric perturbations.
The growth rates calculated for \nirvana{} and polar \flash{} as a function
of time agree within about 10\%.
The results of the linear analysis are consistent with the absence of rapidly
growing vortices near the edge of the gap in our Cartesian-grid PPM
simulations, which is thus not due to artificial numerical damping of
unstable modes.
This type of code does not produce the necessary steepness of
the surface density profile and
does not support growing non-axisymmetric
perturbations.

Both our numerical schemes have very low numerical
diffusion, which is estimated in Appendix~\ref{ap:viscosity}.
Runs with an explicit Navier-Stokes viscosity
were also performed.
The unstable modes in the outer disk calculated for
$\nu=10^{-6}$ have growth rates $\sim0.25\Omega_{\mathrm{K}}$
at 100 orbits, which are 20\% smaller than in
the inviscid calculations.

We speculate that
the Cartesian-grid implementation may be more diffusive for this problem than
polar geometry (in which the unperturbed Keplerian disk flows along the mesh
structure) hence damping the growth of Rossby waves. The linear theory predicts that unstable modes will be present
in the Cartesian simulations after the gap is sufficiently deep
but these modes have smaller growth rate than those obtained from \nirvana{} and \flash{} polar
simulations.
The growth rate for the inner edge of the gap
is larger than it is at the outer gap edge.
This may be artificial because the inner boundary is closer
to the planet position that the outer boundary is to the outer gap edge.

We observe a correlation
 between the growth rate
of the unstable modes in the linear analysis and 
the size of the peaks in the power spectrum of the gravitational torque
on the planet by the disk. This correlation
suggests that vortices in protoplanetary disks can form
close to the gap, produced by an embedded giant planet,
from the collapse of Rossby waves.
Vortices may grow
and be sustained for long timescales by
interaction with the planetary wake
\citep{2003ApJ...596L..91K,2005ApJ...624.1003L}.
The two-dimensional approximation for the disk flow
is anticipated to give qualitatively correct results
although a three-dimensional analysis is needed
to understand heat dissipation in the vertical direction
and refraction effects in radially propagating waves
\citep{1990ApJ...364..326L}.
An important restriction of our simulations
is that the planet is kept on a fixed circular
orbit.
It would be of interest to study how the vortices
rotating along the edge of the gap
affect the migration rate of a freely
moving protoplanet embedded in a 3-dimensional disk.

In summary, the linear analysis confirms that Rossby waves
are formed in a thin protoplanetary disk
with a giant planet with mass ratio
between
$\mu = 10^{-4}$--$10^{-3}$,
within tens of orbital periods.
The unstable modes with larger growth
rates generate a non-axisymmetric perturbation,
with mode number $m=4$--$6$, which breaks into
vortices in the nonlinear regime
producing a non-axisymmetric density distribution.
At the time when the  growth rate becomes positive, small depressions
in vortensity appear along the gap.
These results do not depend on resolution or
boundary conditions.
Simulations with an initial gap also generate vortices
close to the edge of the gap.
The growth rates estimated from the linear theory
agree with the growth rates at late times in simulations
with an initially flat density distribution.
A protoplanetary disk with a giant planet becomes
populated with vortices and spiral shocks
that can efficiently transport angular momentum.
This effect can be important in disks that are
not sufficiently ionized to sustain turbulence
via the MRI instability.
We conclude that vorticity generation in
protoplanetary disks with an embedded giant planet
is a robust mechanism that
can lead to planet formation and
radial transfer of angular momentum.

\begin{acknowledgements}
MdVB was supported by a SAO predoctoral fellowship and a NOT/IAC scholarship.
GD acknowledges support from the NASA Postdoctoral Program.
We thank Artur Gawryszczak for providing his numerical code 
and enlightening discussions.
The support of the RTN
``Planets'' funded by the European Commission under agreement No.
HPRN-CT-2002-0308 is acknowledged during the course of this project.
The \flash\ code used in this work is developed in part by the U.S. Department of
Energy under Grant No. B523820 to the Center for Astrophysical Thermonuclear
Flashes at the University of Chicago.
Some of the calculations reported here were performed on Columbia, operated by
NASA Advanced Supercomputing Division, at NASA Ames Research Center. 
We thank the anonymous referee for comments that improved the manuscript.
\end{acknowledgements}

\bibliographystyle{aa}
\bibliography{references}

\begin{appendix}
\section{Calibration of numerical viscosity}
\label{ap:viscosity}
The calculations presented in this paper are based on inviscid
Euler equations. However, even though neither physical nor 
artificial viscosity terms enter these equations, each numerical
scheme has some intrinsic diffusivity that can be interpreted as
a \textit{numerical} viscosity, $\nu_{\mathrm{n}}$.

To calibrate the numerical viscosity in each of the hydrodynamics 
codes, we use the numerical setup described in Section~\ref{sec:setup}
with mass ratio $q=0$. Tests are performed at two different resolutions
to check for consistency of the results. 
Numerical viscosity is bound to depend on the flow properties. The values 
reported in this Appendix apply to disks in quiescent conditions
and may therefore represent lower limits for numerical diffusion
in models with uniform initial density and fast gap formation.

In the case of \nirvana{}, a time-averaged measure of the numerical 
diffusivity is obtained over a 50 orbit period, as a function of the
radial position, by analysing the trajectories of 500 tracer (massless)
particles released in the disk. The equation of motion of each particle 
is integrated every hydrodynamics timestep by interpolating the velocity 
field at the particle's location and advancing in time its position 
by means of a second-order Runge-Kutta method. The spatial interpolation 
of the velocity field is also second-order accurate. Hence, trajectories 
are formally second-order accurate in both space and time.

We assume that the viscosity is constant in a radial interval 
$\Delta r=0.1$, which contains about 25 equally-spaced tracer 
particles and is orders of magnitude larger than their diffusion 
length scale. We measure the averaged length travelled by the 
particles in each radial interval, over about 50 orbits
(at $r=1$), and estimate the amount of numerical viscosity
under the assumption that the particle drift velocity is
\begin{equation} 
\left|\frac{\ud r}{\ud t}\right| = \frac {3 \nu_{\mathrm{n}}} {2 r}.
\end{equation}
Experiments executed at resolutions 
$n_\mathrm{r} \times n_{\phi}= 256\times768$ and $512\times1536$
provide very similar results.

The largest numerical diffusion is observed close to the inner grid 
boundary, where $\nu_{\mathrm{n}}\sim 10^{-7}$ (in code units, see 
Section~\ref{sec:setup}) at $r\approx 0.55$ and $\sim 10^{-8}$
at $r\approx 0.65$ . In the radial domain between $r\approx 0.75$ and 
$r\sim 1.75$, $\nu_{\mathrm{n}}$ lies between $\sim 10^{-10}$ and 
$\sim 10^{-9}$. 
In the outer part of the simulated disk, the numerical viscosity is 
comprised between $\sim 10^{-9}$ and $\sim 10^{-8}$.

The numerical viscosity in \flash{} is calibrated
using a 2-dimensional
local patch of a Keplerian disk
with a massless sink hole
in the center of the domain
The sink hole has radius equal to the Roche radius
\begin{equation}
R_{\mathrm{R}} = a \left( \frac{q}{3} \right)^{1/3}
\end{equation}
for a protoplanet with mass ratio $q = 10^{-3}$.
The dimensions of the
shearing box are
$20 R_{\mathrm{R}}\times 20 R_{\mathrm{R}}$.
We use local Cartesian coordinates in the Hill
approximation corotating with
the sink hole. The horizontal axis corresponds to the
radial direction and the vertical axis to the direction
of motion of the flow.
Initially, the unperturbed surface density is uniform
and the velocity of the matter flowing into the 
computational domain has only a vertical component given by
the linearized Keplerian velocity
\begin{equation}
   v_y = - \frac{3}{2} \Omega_{\mathrm{K}} x
\end{equation}
We implement periodic boundaries in the vertical
direction.
Our units in the simulation are the Roche radius, $R_{\mathrm{R}}$,
the initial surface density, $\Sigma_0$, and
the Keplerian angular frequency, $\Omega_{\mathrm{K}}$, at the center of the shearing box.
A cavity is produced at the orbital radius of the
sink hole while the gas in the vicinity of the hole
is pulled into it on the viscous timescale.

The equation of viscous diffusion
for a thin accretion disk
can be obtained
in the asymptotic limit
assuming constant numerical diffusivity
\citep{1981ARA&A..19..137P,1999ApJ...514..344B}.
Using the boundary conditions $\Sigma=0$ at $x=0$,
and $\partial\Sigma / \partial x=0$ at $x= x_{\mathrm{out}}$,
where $x_{\mathrm{out}} = 10 R_{\mathrm{R}}$ is the distance
from the center of the shearing box
to the outer radial boundary,
the surface density distribution
for positive $x$
is given by
\begin{equation}
   \Sigma(x,t) = \Sigma_0 \exp \left(- \frac{3 \pi^2 \nu_{\mathrm{n}} t}{4 x_{\mathrm{out}}^2} \right)
   \sin \left( \frac{\pi x}{2 x_{\mathrm{out}}} \right)
\end{equation}
The kinematic viscosity can be evaluated from the ratio of surface densities
at different times
\begin{equation}
   \nu_{\mathrm{n}} = \frac{4 x_{\mathrm{out}}^2}{3 \pi^2 (t_2-t_1)} \ln\left[\frac{\Sigma(t_1)}{\Sigma(t_2)}\right]
\end{equation}
We estimate values of $\nu_{\mathrm{n}}$ between $10^{-8}$ and $10^{-7}$
using the surface density profiles averaged in the vertical direction
at times close to 50 orbits
and grid resolutions of
$n_\mathrm{x} \times n_{\mathrm{y}} = 320\times320$ and $640\times640$.

\end{appendix}

\end{document}